\documentclass[conference,10pt,twocolumn]{IEEEtran} 

\usepackage{cite} 
\usepackage{graphics}
\usepackage{graphicx}
\usepackage[english]{babel}
\usepackage{amsmath}
\usepackage{amsthm}
\usepackage[table,xcdraw]{xcolor}
\usepackage{subfig}
\usepackage[linesnumbered, ruled]{algorithm2e}
\usepackage[noend]{algpseudocode}
\usepackage[font=footnotesize]{caption}
\usepackage{epsfig}
\usepackage{xfrac}
\usepackage{nicefrac}
\usepackage{psfrag}
\usepackage{times}
\usepackage{balance} 
\usepackage{amssymb}
\usepackage{amsfonts}
\usepackage{bm}
\usepackage[bottom]{footmisc}
\usepackage{mwe}
\usepackage{color}
\usepackage{tabularx}
\usepackage{booktabs,multirow}
\usepackage{verbatim}
\usepackage{hyperref}

\usepackage{mwe}
\usepackage{booktabs,multirow}
\usepackage{tikz}
\usetikzlibrary{patterns,arrows}
\usepackage{physics}
\usepackage{verbatim}
\usepackage{wrapfig}
\usepackage{soul}
\usepackage[left=1.62cm,right=1.62cm,top=1.9cm]{geometry}

\newenvironment{bsmallmatrix}
  {\left[\begin{smallmatrix}}
  {\end{smallmatrix}\right]}

\newcommand{\tred}[1]{\textcolor{red}{#1}}

\SetKwRepeat{Do}{do}{while}
\SetKwInOut{Input}{Input}
\SetKwInOut{Output}{Output}
\SetKw{KwBy}{by}

\begin{document}

\title{\fontsize{21}{24}\selectfont Simple and Effective Augmentation Methods for CSI Based Indoor Localization}

\author{ 
Omer Gokalp Serbetci$^\dag$, Ju-Hyung Lee$^{\dag\ddagger}$, Daoud Burghal$^\dag$,  Andreas F. Molisch$^\dag$ \\

	\small $^\dag$Ming Hsieh Department of Electrical and Computer Engineering, University of Southern California,
	\small Los Angeles, USA\\
    \small $\ddagger$ School of Electrical and Computer Engineering, Korea University, Seoul, Korea \\
	\small serbetci@usc.edu, juhyung.lee@usc.edu, burghal@usc.edu, molisch@usc.edu
	}

\maketitle

\begin{abstract}
    Indoor localization is a challenging task. Compared to outdoor environments where GPS is dominant, there is no robust and almost-universal approach. Recently, machine learning (ML) has emerged as the most promising approach for achieving accurate indoor localization. Nevertheless, its main challenge is requiring large datasets to train the neural networks. The data collection procedure is costly and laborious, requiring extensive measurements and labeling processes for different indoor environments. The situation can be improved by Data Augmentation (DA), a general framework to enlarge the datasets for ML, making ML systems more robust and increasing their generalization capabilities. This paper proposes two simple yet surprisingly effective DA algorithms for channel state information (CSI) based indoor localization motivated by physical considerations. We show that the number of measurements for a given accuracy requirement may be decreased by an order of magnitude. Specifically, we demonstrate the algorithms' effectiveness by experiments conducted with a measured indoor WiFi measurement dataset: 
As little as $10\%$ of the original dataset size is enough to get the same performance as the original dataset. We also showed that if we further augment the dataset with the proposed techniques, test accuracy is improved more than three-fold. 
 
\end{abstract}

\section{Introduction} \label{intro}



 Localization is one of the critical components of wireless systems, either for location-based services (e.g., guiding users to a store) or to improve the efficiency of wireless communications (e.g., as a basis for beam tracking in MIMO systems) \cite{zekavat2019handbook}. While outdoor localization can usually perform trilateration using GNSS (global navigation satellite systems) signals, e.g., GPS, they are ineffective in indoor environments. As a matter of fact, despite 
significant theoretical research and test-bed development, existing methods, such as fingerprinting,  trilateration, and proximity-based localization, have all their challenges, often related to the highly complex environmental structure of indoor spaces and the associated propagation characteristics. In light of the general suitability of Machine Learning (ML) for solving complex tasks, the past 15 years have seen significant interest in using ML for indoor localization, and many papers have been published, see \cite{survey} and references therein. These methods provide robust data-driven solutions and have shown excellent performance with versatile setups.


 The first ML-based indoor localization systems used only the received signal strength indication (RSSI) to train straightforward neural networks. Since then, numerous enhancements have been introduced both in terms of the signal characteristics that are employed (use of full Channel State Information CSI, use of multi-antenna signals) and the neural network structures, where a larger number of layers (deep learning), as well as more advanced structures such as LSTM (long short term memory), have been shown to provide superior results.  


Although ML-based indoor localization is a promising approach, there are several challenges to realizing it in practical systems. Chief among them is the availability of data required to train and update such algorithms, which is still perceived as an Achilles' heel. ML solutions inherently require significant data \cite{imagenet}. However, the data acquisition process from each indoor environment is labor- and time-consuming, requiring specialized equipment that can provide both the data and the ground-truth locations and careful labeling and supervision. 

The challenge of data limitation is widespread in ML. While some fields, such as natural language processing or image recognition, have benefited from an abundance of labeled data, this is not true in many other areas, of which indoor localization is one example. 
One approach to address this issue is data augmentation techniques, which  exploit the invariances found in the dataset. They have been used successfully in various areas of ML. For instance, in image classification, rotating, clipping, or flipping images are widely used augmentation techniques\cite{aug_survey}. 
However, augmentation techniques must be tailored to the specific problem to be effective. 
In this paper, we utilize domain knowledge to develop data augmentation techniques for localization problems. In particular, we propose two methods based on phase shift and amplitude fluctuations based on the physics of wireless transceivers and wireless propagation channels, respectively. 

The first algorithm utilizes the unavoidable phase drifts in wireless transceivers. Since transmitters (Tx) and receivers (Rx) have independent clocks and imperfect synchronization, time-variant phase offsets are unavoidable. As a second method, we exploit the fact that there are fluctuations in the amplifiers during CSI measurements. In this method, we randomly add a certain amplitude to the measurements made from a particular measurement device. The proposed augmentation method provides more data for training and makes the localization system more robust to the inherent nonidealities of transceiver hardware. 

 We will show that these methods are simple to implement yet surprisingly effective. In particular, we evaluate
 the performance of these algorithms using a real-life WiFi CSI dataset and demonstrate that with only $10\%$ of the data, we get similar performances to the original dataset. Moreover, if we use the same amount of original data, the augmentation improves the test accuracy of the localization up to three-fold. 



\textbf{Related Works}.\quad Several papers in the literature use full CSI, and not just RSSI, as the basis for their ML indoor localization \cite{resloc, deepfi, cifi, phasefi}. These papers use CSI in various ways, e.g., channel amplitude only, phase information only, or complex CSI, and employ different ML techniques such as convolutional neural networks (CNN) and residual neural networks. Moreover, \cite{crosscount, ganLoc} use WiFi as their data source, whereas \cite{cellindeep} uses cellular data for localization. For a comprehensive survey, please refer to \cite{survey}. 

Only a few papers deal with data augmentation for localization, and those are, to the best of our knowledge, mostly restricted to RSSI-based systems. Ref. \cite{effecive_augment} proposes noise injection, density estimation, random masking, and Variational Autoencoders (VAE) as an augmentation method in an RSSI-based indoor localization system. Another ML approach \cite{ganLoc} uses generative adversarial networks, with a  tensor completion algorithm as the generator that produces new RSSI fingerprints. Ref. \cite{augmentation} uses an RSSI based system, randomly picking beacon frames and adding the average of the picked signals to the RSSI of the selected beacons to augment the dataset. In a recent work, Ref. \cite{gao2022towards} uses noise injection to the channel frequency responses as an augmentation technique.

\textbf{Main Contributions}.\quad  Unlike the papers mentioned above, our work proposes augmentation algorithms for localization with full CSI (compared to the RSSI in the previous work) based on domain knowledge that goes beyond noise injection. The proposed algorithms are simple yet efficient data augmentation techniques that can be implemented as simple preprocessing steps that are easy to implement. Thus the contribution of this paper can be summarized as follows: 
\begin{itemize}
    \item We provide one algorithm based on phase rotation that emulates the realistic behavior of wireless transceivers (introducing independent random phase shifts at different transceivers) to the CSI measurements. This can be seen as a pure augmentation method that always leads to performance improvement. 
    
    \item We provide a second algorithm that mimics the fluctuations in the amplifiers in the CSI measurements. This is another realistic behavior that can occur in any wireless system. This method requires judicious selection of the fluctuation statistics since too large or too little amplitude fluctuation either hurts or leads to no significant performance increase, respectively.
    
    \item The performance of the proposed algorithms is demonstrated based on a measured dataset. The numerical evaluation is shown in various scenarios, including high-low data regimes and testing generalization capabilities. We emphasize that while the simulation results are based on a particular neural network implementation, the augmentation methods can be applied to any network structure.  
\end{itemize}

\section{Background} \label{background}

\subsection{Data Augmentation}
Data augmentation is a technique widely used in ML to enhance the generalization capabilities of models by artificially expanding the training dataset.
In a data-driven ML approach, such as supervised learning, a dataset $\mathcal{D} = \{(\boldsymbol{x}_i, \boldsymbol{y}_i)\}_{i = 1}^N$ is utilized, where $\boldsymbol{x} \in \mathcal{X}$ represents the input feature and $\boldsymbol{y} \in \mathcal{Y}$ denotes the label. For data point $i$, ($\boldsymbol{x}_i, \boldsymbol{y}_i)$ signifies the input feature-label tuple and $N$ corresponds to the total number of data points. 

The objective of the augmentation is to identify a transformation operator $\mathcal{T}: \mathcal{X} \rightarrow \mathcal{X}$ that remains invariant with respect to the dataset $\mathcal{D}$. 
Formally, for a given mapping $f: \mathcal{X} \rightarrow \mathcal{Y}$, the condition $f(\boldsymbol{x}) = f(\mathcal{T}(\boldsymbol{x}))$ must hold. 
This implies that when the augmentation operator is applied to an input feature $\boldsymbol{x}$, the corresponding mapping in the label space $\boldsymbol{y}$ is preserved. 
Common examples in image classification include operations such as rotation and translation of images. In the context of indoor localization problems, our goal is to discover suitable transformation operators $\mathcal{T}$ and their corresponding effects on the dataset.

\begin{figure}[!t]
    \centering
    \includegraphics[width=\linewidth]{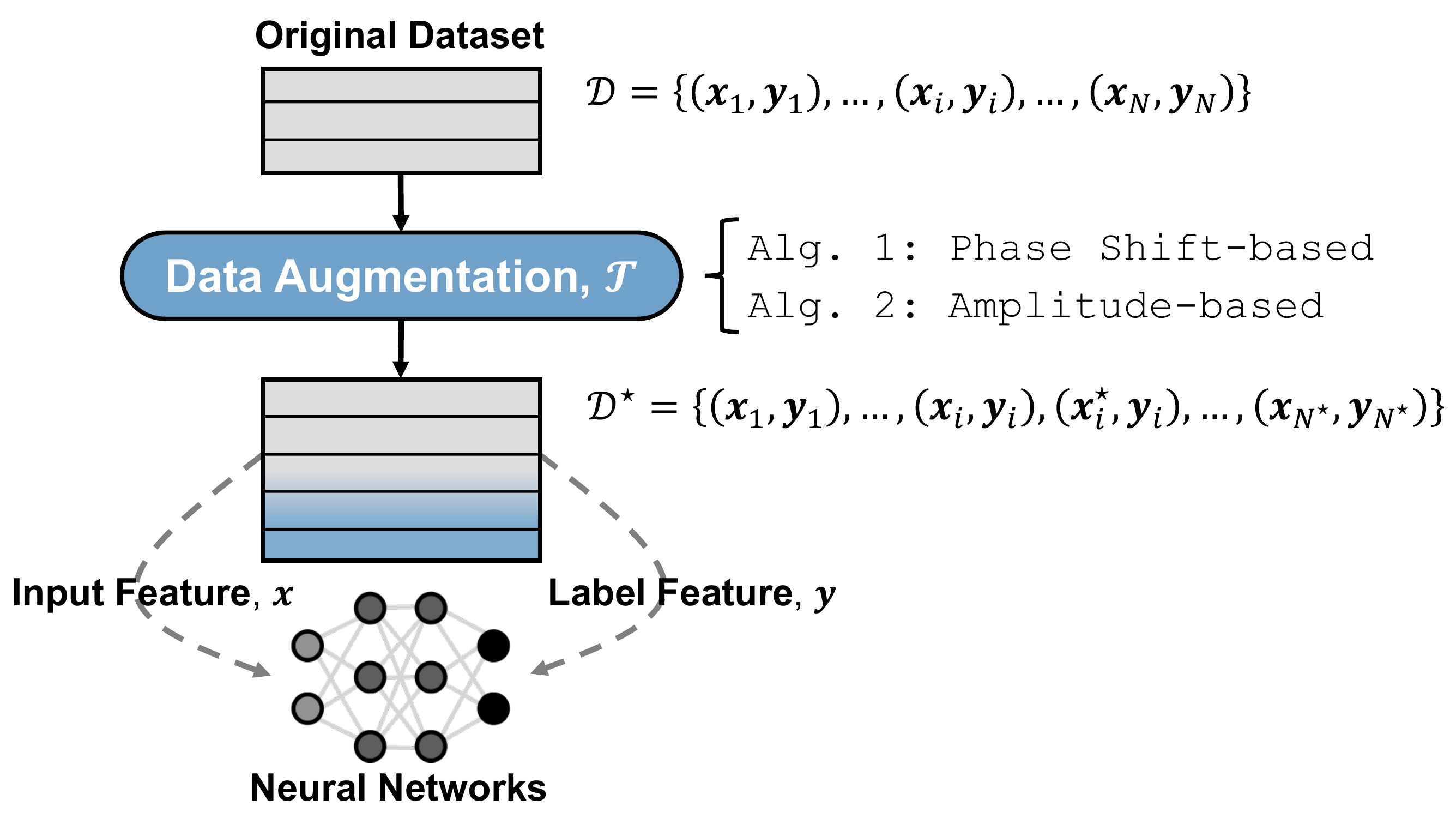}
    \caption{{Illustration of data augmentation process.}}
    \label{aug_figure}
\end{figure}

Fig. (\ref{aug_figure}) shows the data augmentation procedure and how it is connected to the neural network training. Note that data augmentation deals with only training data. Throughout the experiments, we separate the training and test set at the beginning and apply augmentation techniques solely to the training set.

\subsection{Indoor Localization}
We first briefly summarize the methods, features used in the methods, and available ML solutions to the indoor localization problem. The \emph{ML-based} methods considered in this paper use direct coordinates as their labels and use different type input features, such as RSSI or CSI, to find an appropriate matching. 

Input features of the methods mentioned could be RSSI, CSI, CSI amplitude, CSI phase, or the pre-processed features such as Angle of Arrival(AoA) images created from CSI data to feed CNN-based localization algorithms. These inputs can be acquired via wireless technologies such as Bluetooth, WiFi, LTE, etc. \cite{prof_molisch}. These data structures are used with deep learning structures such as CNN and ResNets as supervised learning solutions \cite{survey}. There are also unsupervised and semi-supervised learning solutions that are out of the scope of this paper. Further and more detailed information can be found in the survey paper \cite{survey}.

\section{CSI-based Indoor Localization} \label{sys_model}
In this section, we provide the system model for CSI-based indoor localization, which is the main application of our proposed data augmentation method.

\subsection{System Model}
Assume there are $N_{\mathrm{AP}}$ wireless access points (APs), and each AP has $N_{\mathrm{RX}}$ antennas. 
The system employs orthogonal frequency-division multiplexing (OFDM) with $M$ subcarriers. 
Without loss of generality, here, the localization operates based on uplink transmission.
Let $r_{k, j}$ be the received signal at the $k^{\text{th}}$ AP's $j{\text{th}}$ antenna, where $k \in \{1,2, \dots N_{\mathrm{AP}}\}$, $j \in \{1,2, \dots N_{\mathrm{RX}}\}$. 
Further, let the transmitted signal from position $i$ at subcarrier frequency $f_m$ be $s_i(f_m)$, where $i \in \{1,2, \dots N\}$, and $m \in \{1,2, \dots, M\}$, and $h_{k,j}(f_m)$ be the channel frequency response at the $m^{\text{th}}$ subcarrier with respect to $k^{\text{th}}$ anchor points' $j^{\text{th}}$ antenna. Then, 
\begin{equation}
    r_{k,j}(f_m; i) =  h_{k,j}(f_m;i)s_i(f_m) + w_{k,j}
\end{equation}
where $w_{k,j} \sim \mathcal{CN}(0, \sigma^2_{m, j})$. The noise samples are i.i.d. zero-mean circularly symmetric complex Gaussian samples with variance $\sigma_{k,j}^2$. 
Moreover, the channel response for an environment with $L$ multi-path components (MPC) is:

\begin{equation}
    h_{k, j}(f_m)  \triangleq \sum_{l = 1}^L \alpha_l a_{k,j}(\phi_l, \theta_l, f_m)e^{-j2\pi f_m\tau_l}
\end{equation}
where $a_{k,j}(\phi_l, \theta_l, f_m)$ is the antenna pattern of $j^{\rm th}$ element  with respect to azimuth angle $\phi_l$, elevation angle $\theta_l$, and the MPC has a complex amplitude gain $\alpha_l$. The complex exponential $e^{-j2\pi f_m \tau_l}$ is the characterization of the delay $\tau_l$ in frequency domain. 


\subsection{DL-based Indoor Localization}
We use feedforward neural networks for supervised learning as follows. Let $\mathcal{D} = \{(\boldsymbol{x}_i, \boldsymbol{y}_i)\}_{i = 1}^N$ be the dataset consisting of $N$ measurements (input features). 
For each measurement $\boldsymbol{x}_i$, corresponding label data $\boldsymbol{y}_i$ consist of the information regarding the location of UE, e.g., the coordinate of the UE location and the corresponding fingerprint.
The input feature $\boldsymbol{x}_i \in \mathbb{R}^d$ is a real vector, which consists of the CSI of the measurement, in which the dimension $d$ depends on several factors, such as the number of anchor points, the number of receiver antennas, the number of subcarriers, etc. 
Note that the complex-valued CSI data can be split into real and complex parts and then concatenated for real-value tensors in a neural network. 

In more involved localization algorithms, the input feature $\boldsymbol{x}_i$ could be 2-D data, including, e.g., angle-of-arrival (AOA) information. 
Then, depending on being a fingerprinting or direct coordinate application, the label $\boldsymbol{y}_i \in \mathbb{R}^n$ could be a scalar value for fingerprint index or 2-D coordinate. 
Lastly, the neural network could be trained to handle the localization as a classification or regression problem respectively; in our examples, we will use fully connected feedforward neural networks to solve a regression problem. 

 
We are now ready to formulate the problem of data augmentation for an ML solution to the indoor localization problem. Let $\mathcal{F}$ be the hypothesis class and $\mathcal{A}$ be the algorithm, namely deep learning training procedure when it is fed through the dataset $\mathcal{D}$. Then, let $f: \mathcal{X} \rightarrow \mathcal{Y}$ be the model that is the output of the algorithm $\mathcal{A}$, where $f \in \mathcal{F}$. We assume that the input features are channel response tensors as $\boldsymbol{x} \in \mathcal{X} \subseteq \mathbb{C}^{M \times N_{\mathrm{RX}} \times N_{\mathrm{AP}}}$. The label $\boldsymbol{y} \in \mathcal{Y} \subseteq \mathbb{R}^2 $ is the coordinates of the UEs. 

The problem this paper considers is finding an augmentation operator $\mathcal{T}: \mathcal{X} \rightarrow \mathcal{X}$ such that $\mathcal{R}(f^{\star}) \leq \mathcal{R}(f)$, where $f^{\star}$ is the output of the algorithm $\mathcal{A}$, which is fed by dataset $\mathcal{D}^{\star}$, and $\mathcal{R}(f)$ is the true risk of the model $f$. Dataset $\mathcal{D}^{\star}$ is produced after operator $\mathcal{T}$ is applied on dataset $\mathcal{D}$. 

\section{Data Augmentation Methods for CSI-based Indoor Localization} \label{algos}

This section introduces the data augmentation algorithms for the indoor localization dataset $\mathcal{D}$ consisting of $N$ samples. 

\subsection{Independent Phase Shift-based Data Augmentation}
 The first algorithm is based on the observation that in wireless systems, the clocks of different APs suffer from phase noise and drift that is independent between APs, and independent of the phase of the UE.\footnote{Note that further different phase variation/drift could arise from small scale variation in the environment or variation of the capturing devices, though such changes as observed at the different APs might be correlated.} The fact of such \emph{independent phase shift}  inspires us to generate augmented CSI data by having each AP add an independent phase shift to each recorded measurement signal from the UE. Note, however, that this phase shift is the same over the different subcarriers on one AP since it arises from the same physical source, namely the local oscillator. 
 For example, when looking at only the $k^{\text{th}}$  AP, we generate a random phase $\theta \sim \mathcal{U}[0, 2\pi]$ which then is added to all signals measured by this AP of interest, which corresponds to $M \times N_{\mathrm{RX}}$ different complex channel responses, by multiplying $e^{j\theta}$ with each channel response $h_{k,j}(f_m)$. As a result, a total of $N_{\mathrm{AP}}$ random phases are generated at each augmentation step, and these phases are entered, as complex exponentials, in the associated channel responses. 
 
\begin{algorithm}[!t]
        \KwIn{$\mathcal{D} = \{(\boldsymbol{x}_i, \boldsymbol{y}_i\}_{i = 1}^N$, $N^{\star}$, $N, N_{\mathrm{AP}}, N_{\mathrm{RX}}, M$}
        \KwOut{$\mathcal{D}^{\star} = \{(\boldsymbol{x}_i, {\boldsymbol{y}_i})\}_{i = 1}^{N^{\star}}$}
        $i \gets 1$ \\
        $j \gets 1$\\
        $\mathcal{D}^{\star} \gets \mathcal{D}$\\
        \While{$j + N \leq N^{\star}$}{
          $\boldsymbol{\theta} \gets \mathcal{U}[0,2\pi]^{N_{\mathrm{AP}}}$\\
          ${\boldsymbol{x}_{i}^{\star}} \gets \boldsymbol{x}_i \otimes e^{j\boldsymbol{\theta}}$\\
          $\mathcal{D}^{\star} \gets \mathcal{D}^{\star} \cup (\boldsymbol{x}_{i}^{\star}, \boldsymbol{y}_{i})$\\
          $j \gets j + 1$ \\
          $i \gets i + 1$\\
          \If{$N \leq i$}{
            $i \gets 1$
          }
        }
        \caption{Independent Phase Shift-based Data Augmentation}
        \label{algo1}
\end{algorithm}

Algorithm \ref{algo1} presents the augmentation procedure based on the independent phase shift we discussed above. The operator $\otimes$ below corresponds to the tensor product described earlier. Each element of the random phase vector's $e^{j\boldsymbol{\theta}}$ is multiplied by the corresponding AP's channel response, which is a $M \times N_{\mathrm{RX}}$ matrix.

\subsection{Data Augmentation with Random Amplitude} \label{sec:algo2}
 In the first algorithm we introduced, we tried to mimic the potential phase drift appearing in most wireless systems. We next  propose an augmentation algorithm that emulates the potential amplifier fluctuations, which can result, e.g., from the temperature drift of the amplifiers. To leverage this phenomenon, we uniformly generate an amplitude from the interval $[-P^{\star}, P^{\star}]$ dB for each anchor point, where $P^{\star}$ is a user-defined parameter (alternative statistics of the fluctuations, possibly based on measurements of typical devices, can be used instead). Then, this amplitude is added (on a dB scale) to all measured signals by that anchor point, similar to the procedure in Algorithm \ref{algo1}. Note that by adding these random fluctuations, we do not mimic the fading but the actual fluctuations caused by the measurement device. Furthermore, this fluctuation is also fundamentally different from random noise injection. Algorithm \ref{algo2} provides a detailed description of the procedure. 

\begin{algorithm}[!t]
        \KwIn{$\mathcal{D} = \{(\boldsymbol{x}_i, \boldsymbol{y}_i\}_{i = 1}^N$, $N^{\star}$, $N, N_{\mathrm{AP}}, N_{\mathrm{RX}}, M$}
        \KwOut{$\mathcal{D}^{\star} = \{(\boldsymbol{x}_i, \boldsymbol{y}_i)\}_{i = 1}^{N^{\star}}$}
        $i \gets 1$ \\
        $j \gets 1$\\
        $\mathcal{D}^{\star} \gets \mathcal{D}$\\
        \While{$j + N \leq N^{\star}$}{
          $\boldsymbol{P} \gets \mathcal{U}[0,2\pi]^{N_{\mathrm{AP}}}$\\
          $\boldsymbol{x}_{i}^{\star} \gets \boldsymbol{x}_i \otimes \boldsymbol{P}$\\
          $\mathcal{D}^{\star} \gets \mathcal{D}^{\star} \cup (\boldsymbol{x}_{i}^{\star}, \boldsymbol{y}_{i})$\\
           $j \gets j + 1$ \\
           $i \gets i + 1$\\
           \If{$N \leq i$}{
            $i \gets 1$
          }
        }
        \caption{Random Amplitude-based Data Augmentation}
        \label{algo2}
\end{algorithm}

\section{Numerical Evaluation} \label{num_results}

This section shows the impact of our proposed two data augmentation methods (Algorithms 1 and 2) for indoor localization. For the experiments, a {\em measured} WiFi CSI dataset, \emph{Wireless Indoor Localization Dataset} (WILD) \cite{wild}, is used as the original dataset; the details are elaborated in the following.

\textbf{Evaluation}.\quad
For the main performance metric of the DNN-based indoor localization, mean square error (MSE) is considered, which evaluates the level of misestimation (\textit{e.g.}, the accuracy of localization) in the entire dataset. 
The lower the values of MSE, the better. 
The MSE is computed from the predicted $\hat{f}(\vb* {x}_i)$ and the ground-truth $\vb* y$:
\begin{equation}\label{mse_eq}
    \mathrm{MSE}(\{\vb* {x}_i, \vb* {y}_i\}_{i = 1}^N)  = \frac{1}{N}\sum_{i = 1}^N \lVert \hat{f}(\vb* {x}_i) - \vb* {y}_i\rVert^2_2  
\end{equation}
where $\hat{f}$ is the trained neural network model.

\textbf{WILD Dataset}.\quad
The WILD dataset contains measurements from two different environments separately. The first one is an NLOS environment over 1500 sq. ft. with $N_{\mathrm{AP}}= 4$ APs, and the second one is a 500 sq. ft. LOS environment with $N_{\mathrm{AP}} = 3$ APs, where each AP has $N_{\mathrm{RX}} = 4$ RX antennas. The dataset is based on a WiFi system with $M = 234$ subcarriers. Moreover, $N = 51613$ data points for the NLOS environment and $N = 56395$ for the LOS environment are labeled. The dataset presents the measurements as a complex 4-D tensor, i.e., $N \times M \times N_{\mathrm{RX}} \times N_{\mathrm{AP}}$. The corresponding UE coordinates labels are $(x,y) \in \mathbb{R}^2$.

With the original dataset, we demonstrate the effect of our proposed augmentation method in three different (size of original dataset) scenarios: \textit{i)} \emph{small} ($4$K samples), \textit{ii)} \emph{medium} ($20$K samples), and \textit{iii)} \emph{large} ($40$K samples).   
Here, a small data regime refers to a small dataset size compared to the usual deep learning applications. The medium case aims to show the transition from low to high data regimes. Finally, we are interested in how much performance gains in high data regimes, where we have enough data to perform reasonably.

\begin{figure}[!t]
    \centering
    \subfloat[{LOS} \label{fig:LOS}]{\includegraphics[width=.45\columnwidth]{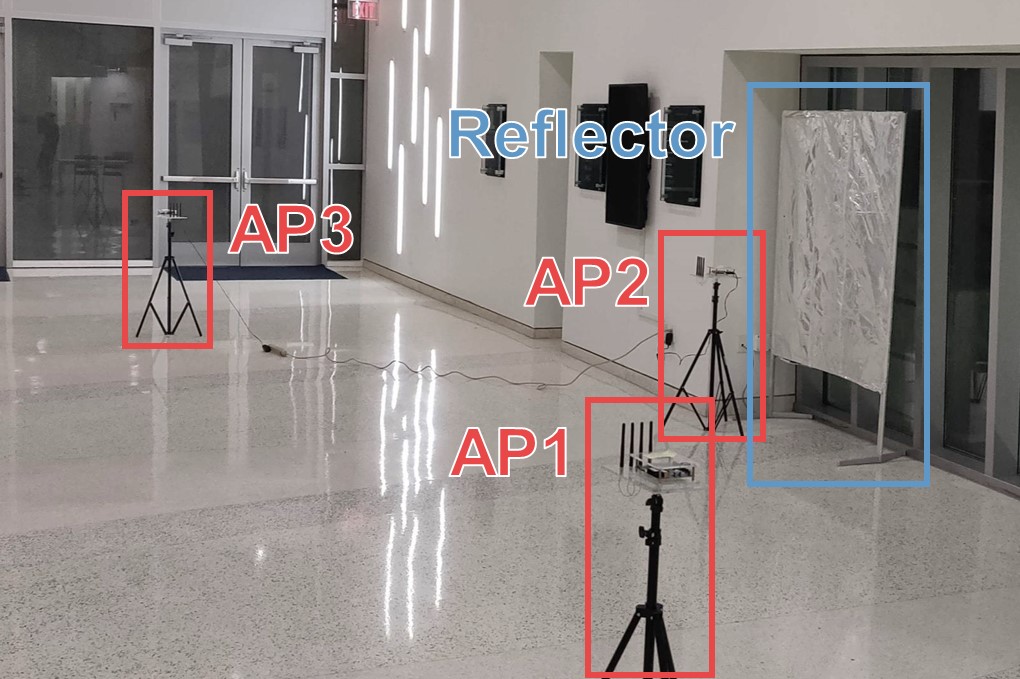}} \
    \subfloat[{NLOS} \label{fig:NLOS}]{\includegraphics[width=.53\columnwidth]{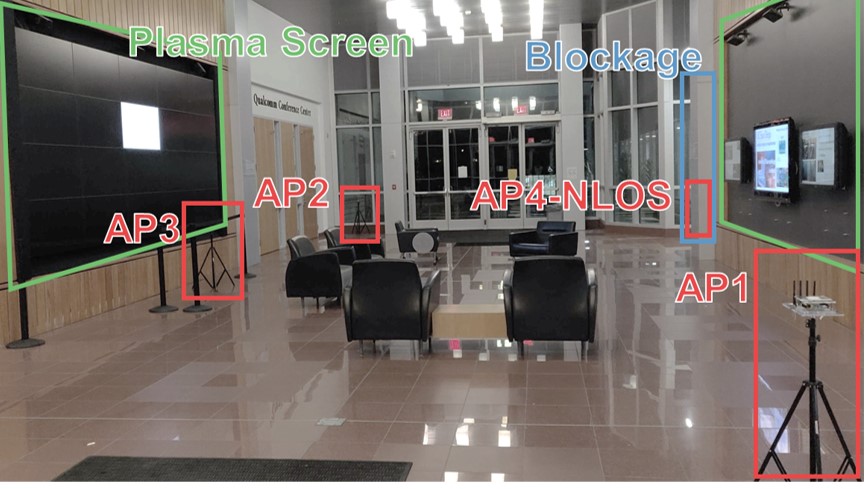}} 
\caption{{Measurement scenario of original dataset (WILD \cite{wild}).}}
    \label{fig:WILD}
\end{figure}

\begin{table}[h!]   
  \centering
  \resizebox{\columnwidth}{!}{\begin{minipage}[t]{0.9\columnwidth}
  \centering
  \begin{tabular} {l l l }
	\toprule[1pt]
	\textbf{Parameter} & \textbf{LOS} & \textbf{NLOS} \\
	\cmidrule(lr){1-1} \cmidrule(lr){2-2} \cmidrule(lr){3-3}
	Scale of Environment & $500$ sq. ft. & $1500$ sq. ft. \\
	\# of AP ($N_{\mathrm{RX}}$) &   4 &   4 \\  
	\# of Antenna ($N_{\mathrm{RX}}$) &   3 & 4 \\  
	\# of Subcarrier  ($M$) & 234 & 234\\  
	\# of Data Points ($N$) & 56395 & 51613 \\  
	\bottomrule[1pt]
\end{tabular}

  \label{table_Paramter}
  \end{minipage}}
  \caption{{Measurement setup/scenario for original dataset (WILD \cite{wild}).}}
\end{table}

\subsection{Impacts of Independent Phase Shift-based Data Augmentation (Algorithm \ref{algo1})}
To evaluate the performance of the independent phase shift as an augmentation method, we used a feedforward neural network with three hidden layers, each consisting of 256 neurons with ReLU activation function and trained for 300 epochs. The augmentation method is applied to NLOS and LOS datasets separately.

Each numerical column of Table \ref{algo_1_small}, Table \ref{algo_1_medium}, and Table \ref{algo_1_large} gives the test set performance with respect to MSE in meters. \emph{No Augmentation} refers to training with the original data only, which in the low-data case, we train with only 4000 samples. The other columns refer to training with a dataset augmented by the given multiple compared to the original size. For example, a low data regime with $\times 6$ means training with 24000 samples, where 20000 samples are generated with the augmentation algorithm and 4000 samples are the original data.

In all cases, data augmentation significantly improves performance. First, compare the accuracy improvement for a given number of measured labeled data; in this case, augmentation provides up to 3.25 times better MSE. Of particular practical importance is the reduction of the MSE in the NLOS case in the low-data case, where augmentation allows to realize reasonable ($1.5$ m) accuracy instead of more than $5$ m without augmentation. Moreover, we improve the localization performance even when the measured data set is very large, reducing the MSE from $0.8$ to $0.4$ m in NLOS. 

Another practically relevant question is: \textit{How much can we reduce the size of the {\em measured and labeled} training set without losing accuracy?} We thus performed an additional experiment in which we augmented the low data regime to create a set as large as the high data case, in other words, $\times 10$ of the initial dataset size. In the NLOS case, we get 0.823344 m, and in the LOS case, 0.316802 m MSE, similar to \emph{No Augmentation} cases of high data regimes. These results show that it is possible to achieve the same performance but with only 10\% of the required measurement/labeling effort (remember that the high data regime has 40000 samples, whereas we have only 4000 samples in the low data regime!). 

\begin{table}[t!]
    \centering
    \caption{{Impact of Algorithm 1 for \emph{small} original dataset. MSE score comparison  with respect to augmentation size.}}
    \resizebox{1.\columnwidth}{!}{
        \begin{minipage}[h]{.9\columnwidth}
        \centering
       \label{algo_1_small}
        \begin{tabularx}{1\linewidth}{l c c }
\toprule[1pt]
\textit{Size of Training Set} & \textit{MSE (LOS Env.)} & \textit{MSE (NLOS Env.)} \\
\cmidrule(lr){1-1} \cmidrule(lr){2-2} \cmidrule(lr){3-3}

w/o Aug. ($1$X) & $0.994347 $\; \
& $5.204818$\; 
\\ 

w Aug. ($2$X) & $0.599574$\; 
& $3.089172$\; 
\\ 

w Aug. ($3$X) & $0.653126$\;
& $1.608795$\; 
\\ 

w Aug. ($4$X) & $ 0.624667$\; 
& $2.067157$\; 
\\ 

w Aug. ($5$X) & \textbf{$0.340235$}\;  
& $\textbf{1.349240}$\; 
\\ 

w Aug. ($6$X) & \textbf{$\textbf{0.307535}$}\; 
& $1.631480$\;
\\

\bottomrule[1pt]
\end{tabularx}

        \end{minipage}
    }
\end{table}

\begin{table}[t!]
    \centering
    \caption{{Impact of Algorithm 1 for \emph{medium} sized original dataset. MSE score comparison  with respect to augmentation size.}}
    \resizebox{1.\columnwidth}{!}{
        \begin{minipage}[h]{.9\columnwidth}
        \centering
        \label{algo_1_medium}
        \begin{tabularx}{1\linewidth}{l c c }
\toprule[1pt]
\textit{Size of Training Set} & \textit{MSE (LOS Env.)} & \textit{MSE (NLOS Env.)} \\
\cmidrule(lr){1-1} \cmidrule(lr){2-2} \cmidrule(lr){3-3}

w/o Aug. ($1$X) & $0.402769$\; 
& $1.238651$\; 
\\ 

w Aug. ($2$X) & $0.316508$\; 
& $ 1.067086$\; 
\\ 

w Aug. ($3$X) & $0.183591$\; 
& $0.804732 $\; 
\\ 

w Aug. ($4$X) & $ 0.152524$\; 
& $0.777409$\; 
\\ 

w Aug. ($5$X) & $0.130289$\; 
& $0.613748$\; 
\\ 

w Aug. ($6$X) & \textbf{$\textbf{0.123952}$}\; 
&  \textbf{$\textbf{0.523407}$}\; 
\\ 

\bottomrule[1pt]
\end{tabularx}

        \end{minipage}
    }
\end{table}

\begin{table}[t!]
    \centering
    \caption{{Impact of Algorithm 1 for \emph{large} original dataset. MSE score comparison  with respect to augmentation size.}}
    \resizebox{1.\columnwidth}{!}{
        \begin{minipage}[h]{.9\columnwidth}
        \centering
        \label{algo_1_large}
        \begin{tabularx}{1\linewidth}{l c c }
\toprule[1pt]
\textit{Size of Training Set} & \textit{MSE (LOS Env.)} & \textit{MSE (NLOS Env.)} \\
\cmidrule(lr){1-1} \cmidrule(lr){2-2} \cmidrule(lr){3-3}

w/o Aug. ($1$X) & $0.252251$\; 
& $0.824396$\; 
\\ 

w Aug. ($2$X) & $0.149603$\; 
& $0.635322$\; 
\\ 

w Aug. ($3$X) & $0.130312$\; 
& $0.574232 $\; 
\\ 

w Aug. ($4$X) & $ 0.112951$\; 
& $0.451495$\; 
\\ 

w Aug. ($5$X) & $0.099951$\; 
& $0.427812$\; 
\\ 

w Aug. ($6$X) & $\textbf{0.093153}$\; 
& $\textbf{0.395018}$\; 
\\

\bottomrule[1pt]
\end{tabularx}

        \end{minipage}
        \vspace{-3em}
    }
\end{table}

\subsection{Impacts of Random Amplitude-based Data Augmentation (Algorithm \ref{algo2})}
For the experiments with the random-amplitude augmentation, we have used 75 to 150 epochs, and 
the neural network architecture is kept the same as the previous experiment. We used $P^{\star} = 1.5$ dB in the small and medium data cases, but $P^{\star} = 0.75$ dB in the large-data case (see below for motivation).

We first notice that Algorithm \ref{algo2} does not provide performance improvement for the large data set and can actually degrade performance due to overfitting in the rich data set. The problem would be even more pronounced if we used $P^{\star}=1.5$ in the large-data case, so we adopted the smaller value of $0.75$. 

However, Algorithm \ref{algo2} provides significant performance improvements for medium and small datasets. Tables \ref{algo_2_small} and \ref{algo_2_medium} demonstrate the significant performance improvements: up to 2 and 1.7 times reduction of the MSE in the NLOS and LOS in the small dataset case, respectively.


\begin{table}[h!]
    \centering
    \caption{{Impact of Algorithm 2 for \emph{small} original dataset. MSE score comparison  with respect to augmentation size.}}
    \resizebox{1.\columnwidth}{!}{
        \begin{minipage}[h]{.9\columnwidth}
        \centering
        \label{algo_2_small}
        \begin{tabularx}{1\linewidth}{l c c }
\toprule[1pt]
\textit{Size of Training Set} & \textit{MSE (LOS Env.)} & \textit{MSE (NLOS Env.)} \\
\cmidrule(lr){1-1} \cmidrule(lr){2-2} \cmidrule(lr){3-3}

w/o Aug. ($1$X) & $1.41290116 $\; 
& $5.72591114$\; 
\\ 

w Aug. ($2$X) & $1.35329723 $\; 
& $  3.97055149 $\; 
\\ 

w Aug. ($3$X) & $1.04413772$\; 
& $3.59251714 $\; 
\\ 

w Aug. ($4$X) & $  0.99395692$\; 
& $3.81182671  $\; 
\\ 

w Aug. ($5$X) & $ 0.87801009 $\; 
& $3.08143544 $\; 
\\ 

w Aug. ($6$X) & $\textbf{0.83151478}$\; 
&  $\textbf{2.85603070} $\; 
\\ 

\bottomrule[1pt]
\end{tabularx}
        \end{minipage}
    }
\end{table}
\begin{table}[h!]
    \centering
    \caption{{Impact of Algorithm 2 for \emph{medium} sized original dataset. MSE score comparison  with respect to augmentation size.}}
    \resizebox{1.\columnwidth}{!}{
        \begin{minipage}[h]{.9\columnwidth}
        \centering
        \label{algo_2_medium}
        \begin{tabularx}{1\linewidth}{l c c }
\toprule[1pt]
\textit{Size of Training Set} & \textit{MSE (LOS Env.)} & \textit{MSE (NLOS Env.)} \\

\cmidrule(lr){1-1} \cmidrule(lr){2-2} \cmidrule(lr){3-3}

w/o Aug. ($1$X) & $0.51810563 $\; 
& $1.41850710  $\; 
\\ 

w Aug. ($2$X) & $0.40300253$\; 
& $ 1.57022965 $\; 
\\ 

w Aug. ($3$X) & $ 0.38420656$\; 
& $1.62614357  $\; 
\\ 

w Aug. ($4$X) & $\textbf{0.36034465}$\; 
& $\textbf{1.23139620}$\; 
\\ 

w Aug. ($5$X) & $  0.41608047 $\; 
& $ 1.52679873 $\; 
\\ 

w Aug. ($6$X) & $0.39935032$\; 
&  $1.34331584$\; 
\\ 

\bottomrule[1pt]
\end{tabularx}

        \end{minipage}
    }
\end{table}
\begin{table}[h!]
    \centering
    \caption{{Impact of Algorithm 2 for \emph{large} original dataset. MSE score comparison  with respect to augmentation size.}}
    \resizebox{1.\columnwidth}{!}{
        \begin{minipage}[h]{.9\columnwidth}
        \centering
        \label{algo_2_large}
        \begin{tabularx}{1\linewidth}{l c c }
\toprule[1pt]
\textit{Size of Training Set} & \textit{MSE (LOS Env.)} & \textit{MSE (NLOS Env.)} \\
\cmidrule(lr){1-1} \cmidrule(lr){2-2} \cmidrule(lr){3-3}

w/o Aug. ($1$X) & $0.24627791 $\; 
& $\textbf{0.80115551}$\; 
\\ 

w Aug. ($2$X) & $0.22379075$\; 
& $ 1.18527663 $\; 
\\ 

w Aug. ($3$X) & $ \textbf{0.21959668}$\; 
& $0.81640501   $\; 
\\ 

w Aug. ($4$X) & $0.23696624$\; 
& $ 0.98888855  $\; 
\\ 

w Aug. ($5$X) & $ 0.25395477 $\; 
& $1.08478308$\; 
\\ 

w Aug. ($6$X) & $0.25659689$\; 
&  $1.02641153$\; 
\\ 

\bottomrule[1pt]
\end{tabularx}

        \end{minipage}
    }
\end{table}
We also implement random noise injection as a baseline, adding each data point zero mean unit variance circular symmetric complex Gaussian random variable realization. However, the MSE reduction is between $0$ and $50\%$, i.e., significantly less than the factor-3 reduction our method achieves. Finally, data augmentation is significantly more helpful in small dataset regimes than large dataset regimes. Test set performance can suffer from overfitting when data augmentation is applied to large datasets already rich enough to perform well. Table \ref{algo_2_medium} and Table \ref{algo_2_large} show this bad effect of data augmentation over large datasets.

\section{Conclusion} \label{conclusion}
    In this work, we introduced two novel and simple yet effective data augmentation techniques for deep-learning-based indoor localization using CSI. Based on the physical phenomena of the localization transceivers, namely phase drift and amplifier fluctuations, the augmentation methods reduce the required amount of measured and labeled data, thus reducing the cost and labor for establishing suitable training sets. Results from a real-world WiFi dataset showed that augmentation is very effective, especially in small and medium-sized datasets. The random-amplitude approach is effective for small data sets but might cause overfitting in some scenarios where we already have many training samples. On the other hand, the random-phase approach provides performance improvements in {\em all} cases and may thus be the preferable solution. 

According to the experiments, our method allows the same performance by using a data set of only $10\%$ of the original data. Conversely, MSE can be improved by up to a factor of 3 by augmenting.

\vspace{-.6em}

\section*{ACKNOWLEDGEMENT} 
    Authors greatly appreciate the constructive discussions with Mike Neumann and Dr. Lei Chu. The work was financially supported by NSF grant 2003164.
\vspace{-.6em}

\bibliographystyle{IEEEtran}
\bibliography{bibli}

\end{document}